# Assessment method for photo-induced waveguides

**Mathieu Chauvet[1-2)], Guoyuan Fu[1)], and Gregory Salamo[1)]**

[1)]University of Arkansas, Physics department, 72701Fayetteville, AR, USA
[2)]University of Franche-Comté, FEMTO-ST Institute, 25000 Besancon, France
*mathieu.chauvet@univ-fcomte.fr*

**Abstract:** A method to assess the guiding characteristics of waveguides formed in real-time is proposed and evaluated. It is based on the analysis of the time dependent light distribution observed at the exit face of the waveguide while progressively altering its index profile and when excited by a large probe beam at the input. A beam propagation method is used to model the observed dynamics. The technique is applied to retrieved properties of soliton-induced waveguides.

**Introduction**

Optically induced waveguides constitutes a challenging research subject whose motivation is the fabrication of complex optical circuits in 2-D or even 3-D if realized deep inside crystals. These waveguides can be formed in numerous medium such as polymers [1-3], liquid crystals [4], glass [1, 5, 6] or photorefractive crystals [7-9]. Realization of these structures would benefit from a technique allowing real time evaluation during the photo-induction process. This would provide a tool to characterize the induction process and would ultimately give an accurate control of the characteristics of the fabricated waveguides. Different techniques such as refractive index mapping using interferometry [8] or higher order mode excitation [7] are already employed for this purpose but more versatile characterization tools that take advantage of the specificity of photoinduced waveguides would be helpful.

In this paper we propose a simple method to assess the properties of photo-induced waveguides. The method relies on the spatio-temporal analysis of light distribution at the exit face of a waveguide while its index profile is gradually changing either during the induction process or when the waveguide is gradually erased. After briefly presenting the theoretical basis, the technique is applied to analyze waveguides formed by photorefractive spatial soliton-like beams in a strontium barium niobate (SBN) crystal.

**Principle**

When coherent light is launched at the entrance face of a waveguide it redistributes among guided and leaky modes. As it propagates, it gives rise to a complex light distribution due to the coherent superposition of these multiple independent waves each traveling with a different velocity. This phenomenon is commonly referred as mode beating. For given launching conditions the light distribution along propagation and the waveguide properties are closely interrelated. Knowledge of light distribution as it travels can thus lead to waveguide characteristics. However, in most case, accurate observation of light inside the waveguide is unachievable. As an alternative we propose to analyze light distribution as a function of time at the exit face of the waveguide while the index structure gradually evolves. This techniques is especially suitable for photo-induced waveguides since measurements can be performed in real time during waveguide formation or afterwards if waveguide erasure is possible. In both case the guiding properties are inferred from the observed dynamic.

More specifically, we propose to probe the photo-induced structure with a plane-wave like beam in order to excite not only guided modes but also to launch light in the waveguide surrounding which, as shown later, improves the sensitivity of the technique for singlemode waveguides. Note that the overlap integral between the excitation beam and odd modes being null only even modes are excited. The movie composed of the sequence of images observed at the exit face of the changing waveguide is compared with the movie from numerical calculations based on a beam propagation method (BPM) using the split step Fourier technique [10].

**Validation of the technique**

To validate the technique we consider slab waveguides induced by 1-D bright photorefractive spatial screening soliton formed in SBN:75 crystals [7, 11]. The dynamic observed at the exit face of the waveguide is registered while the waveguide is decaying instead of during its formation to avoid any influence of probe beams on the writing process.

According to photorefractive time-dependent soliton 1-D theory [12] the refractive index change produced by a photorefractive screening soliton whose profile is $I(x)$ is given by:

$$\Delta n(x) = -0.5 \ n_0^3 \ r_{eff} \frac{E_0}{I(x)+I_d}(I_d + I(x)\exp(-\frac{I(x)}{I_d T_d}t))) \quad (1)$$

Where $r_{eff}$ is the effective electro-optic coefficient, $E_0$ is the external applied field, $I_d$ is the equivalent dark irradiance, $T_d$ is the dielectric response time of the crystal in the dark and t is the photo-induction duration time. Evidently in order to obtain the refractive index profile, the soliton intensity profile $I(x)$ has to be determined in self-consistency with the non-linear Schrödinger propagation equation as in refs. [11, 12]. To form a steady-state soliton [13] the induction time has to be longer than $T_d I_d/I_{max}$ while a soliton in quasi-steady-state regime [14, 15] is obtained for an induction time $t_0 = 2I_{max}T_d/I_d$ [12, 16]. Solitons in quasi-steady-state regime have been used in this study because less stringent parameters control is needed for their creation compare to steady-state solitons. Especially, no added background illumination is necessary and $I_d$ is only due to thermal generation of free carriers. In addition because soliton beam intensity is much higher than $I_d$ the soliton width is intensity independent [ref]. Once the soliton is formed both the applied field and the soliton beam are turned off. When subsequently probed by a large probe beam the waveguide depth is assumed to decay exponentially in our model due to gradual erasure. Note however that for the proposed technique precise knowledge of the waveguide decay constant is not required but on the contrary its value can be evaluated experimentally.

A 1 cm$^3$ poled SBN:75 cubic-shaped sample is used to performed the experiments. The first step is to create a 1-D photorefractive soliton using ordinary light at 514 nm. The experiment consists in focusing the beam from a laser using a cylindrical lens into a 15 µm FWHM wide stripe at the entrance face of the SBN crystal. The stripe propagates perpendicular to the crystal c-axis while the external dc electric field $E_0$ is applied along c-axis. When the applied electric field is set to 3kV/cm the initial beam diffraction is gradually compensated as the photorefractive effect takes place. At the light intensity used for the experiment the time $t_0$ to form a quasi-steady-state soliton is about 20 min. Then the soliton and the dc applied field are both turned off and an ordinary polarized 300 µm diameter probe beam is launched normal to the entrance face of the photoinduced waveguide. The image of the light distribution at the exit face is registered at regular interval using a CCD camera linked to an acquisition system.

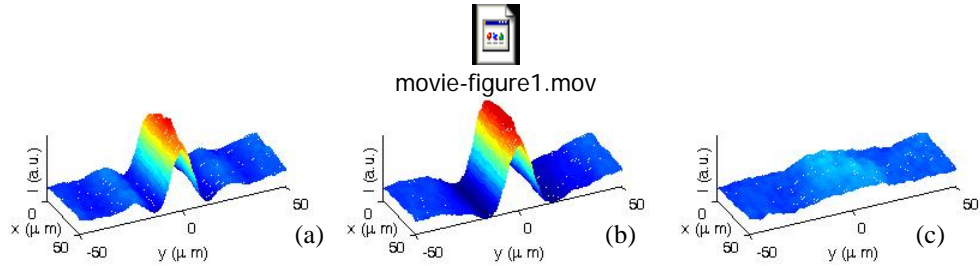

Fig. 1. Observed light distribution at the exit face of a decaying waveguide initially formed by a quasi-steady-state soliton in the initial stage (a) and at two subsequent characteristic times (b, c). Movie of the entire process corresponding to 40min observation is attached (size : 1.35 Mb).

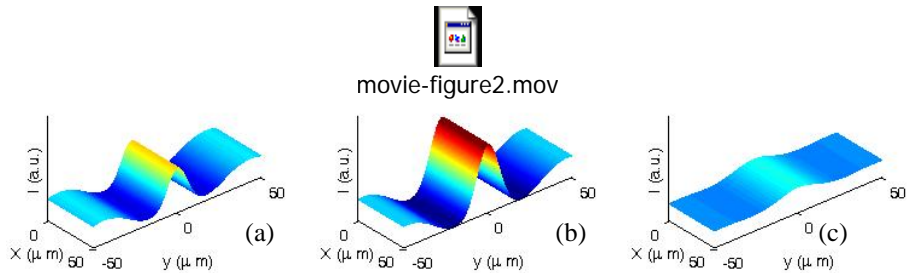

Fig. 2. Calculated dynamic of light distribution at the exit face of a decaying soliton-induced in the initial stage (a) and at two subsequent characteristic times (b, c). Movie of the entire process is attached (size : 640 kB).

Unless fixed [17] PR waveguides usually decay due to thermal excitation or can even be erased faster if illuminated with a beam at a sensitive wavelength. In our experiments, the probe beam is both used to sense the waveguide and to gradually erase it. The constant time associated with the decay is inversely proportional to the illumination as usual for photorefractive effect. Accordingly, the probe intensity is adjusted so that the erasure process can easily be observed in real time and images can be capture at regular intervals.

The first assessed waveguide is formed with a 20 min induction time which corresponds to a soliton in quasi-steady-state regime for our experimental parameters. Evolution of the intensity distribution at the exit face of the waveguide as the photo-induced structure decays is depicted in figure 1. At start, presence of a peak intensity due to light trapping in the high index region of the medium can be clearly seen (Fig. 1a). As the waveguide weakens a remarkable feature we notice is that the contrast between the darkest and the brightest regions increases (Fig. 1b). Then the intensity distribution smoothens as the refractive index distribution becomes flat (Fig. 1c). The entire process which lasts 40 min is summarized in the movie from figure 1. To model this dynamic we consider that the initial index profile is given by Equ. (1) with the inducing spatial soliton being described by the experimental parameters (15 μm FWHM, $E_0$ = 3 kV/cm) with t = $t_0$ and the only free parameters is the electro-optic coefficient. Assuming an exponential decay time constant for the waveguide erasure we obtain the calculated behavior presented in figure 2 when the electro-optic coefficient is set to 40 pm/V for best agreement with experiments. The overall dynamics calculated numerically agrees well with the experimental observation. In particular the increased light modulation that appears after the waveguide has weakened is clearly reproduced by the simulation. The value of the electro-optic coefficient $r_{13}$ used to fit the experimental results is lower than the one (70 Pm/V) usually reported in the literature [ref] which can be attributed to a poorly poled SBN sample. The observed non monotonic decay of the trapped light as the waveguide declines may seem surprising at first especially if we keep in mind that the waveguide is singlemode. Even if waveguide formed by photorefractive solitons are not always singlemode it is indeed the case for steady-state regime soliton with a ratio $I_{max}/I_d$ close to 2 [7] and since a quasi-steady-state soliton has identical characteristic that the latter mentioned soliton its singlemode character is also expected. We then cannot attribute the non monotonous light distribution observed at the exit face to coherent addition of multiple guided modes. The effect is instead due to the coherent overlap between the guided fundamental mode and light in the vicinity of the waveguide each having a different traveling speed consequently giving rise to beating. Although the overall behavior as a function of time depicted by theory agrees with the experiment, a simple exponential decay does not fit accurately the time evolution of the observations. The primarily reason is that a varying light distribution is responsible in practice for the waveguide erasure contrary to the hypothesis of a uniform illumination. For instance, we observe that, in the initial stage, alteration of the observed light distribution shown in Fig. 1 is happening slower compared with predictions (Fig. 2). An obvious reason is that faint light is present initially on both side of the waveguide and as a consequence the photorefractive charges that mainly reside in these regions are not efficiently optical excited. The waveguide decay occurs at a reduced speed until some light appears in these regions. A more accurate description of the time dependence of the process would thus implies to take into account the temporal variations of light distribution on the waveguide decay. This refinement is not necessary if the goal is to assess the geometrical and optical properties of the initial waveguide instead of determining the dynamic of the erasure. We would like to point out that no partial reconstruction of the waveguide is possible under the influence of the probe beam because of the absence of applied electric field, spatial charges tends to redistribute uniformly which induces the waveguide gradual disappearance.

movie-figure3.mov

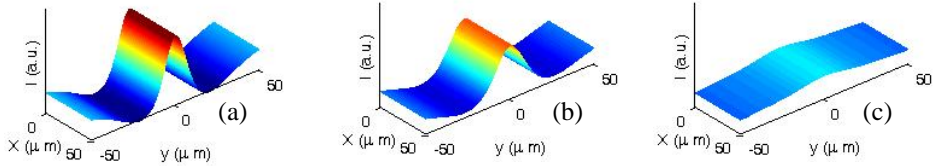

Fig. 3. Calculated dynamic of light distribution at the exit face of a decaying waveguide initially created by a partially formed soliton in the initial stage (a) and at two subsequent characteristic times (b, c). Movie of the entire process is attached (size : 548 kb).

movie-figure4.mov

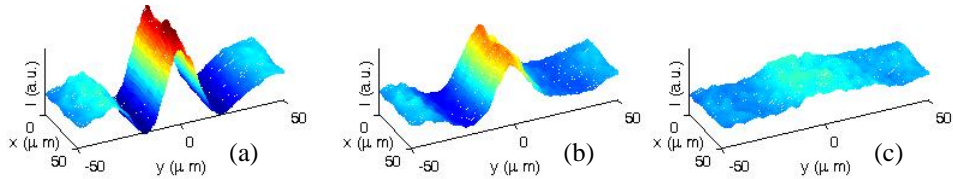

Fig. 4. Observed light distribution at the exit face of a decaying waveguide initially created by a partially formed soliton in the initial stage (a) and at two subsequent characteristic times (b, c). Movie of the entire process corresponding to 20min observation is attached (size : 965 kb).

The assessment technique is then performed on a waveguide induced by a partially formed soliton. In this case, the process to form a soliton is repeated identically than for the previous experiment except that the illumination time is limited to 5 minutes. Because of a shorter photoinduction time the soliton regime is not fully reached and the waveguide is weaker. The beam diffraction is however strongly reduced and the induced index transverse profile can be considered fairly homogeneous along the 1cm crystal length. The refractive index profile is expected to be correctly described by Equ. (1) because of the closeness to soliton regime. When probed by a large beam the theoretical predicted dynamical observation is shown in figure 3 for an induction time set to $t_0/4$ while the other parameters are kept identical as the calculation from figure 2. This theoretical result fits remarkably well the experimental observation from figure 4. At start, we observed that the light focused in the waveguide is separated from the surrounding probe beam by a pronounced dark region. Afterward this contrast decreases monotonically as a function of time. These two observations along with a initial wider focused beam constitute the main differences with the behavior shown in figure 2. It also confirms that the technique can be used to recover waveguides properties even for singlemode waveguides.

movie-figure5.mov

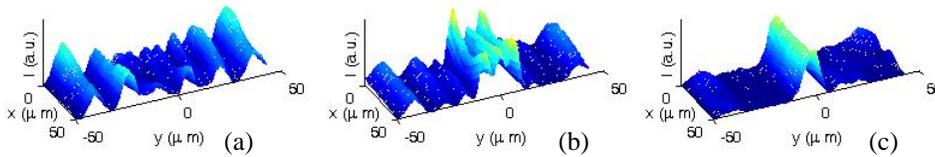

Fig. 5. Observed light distribution at the exit face of a decaying waveguide initially created by a quasi-steady-state soliton and probed by an extraordinary polarized beam, in the initial stage (a) and at two subsequent characteristic times (b, c). Movie of the entire process corresponding to 28min observation is attached (2.2 MB).

Finally, another soliton-induced waveguide is then formed in the crystal in order to apply the technique to a multimode waveguide. The waveguide is still produced with an ordinary polarized quasi-steady-state with experimental parameters identical than for the results from figure 1 except we consider the index change corresponding to an extraordinary polarized beam. For that reason, the waveguide is probed with a beam polarized along the crystal c-axis thus taking advantage of the strong $r_{33}$ electro-optic coefficient of SBN. The dynamics observed at the exit face is presented in Fig. 5. Dramatic changes are clearly exposes in the movie. The complex mode beating reveals that the waveguide supports 3 guided modes. It also exposes at some instant an almost homogeneous intensity despite the presence of a strong waveguide as for example in Fig. 5a. This makes clear that a simple observation of light distribution at the exit face of a fixed waveguide is not sufficient to deduce its guiding properties. In the present case observation limited to the initial condition would lead to the wrong conclusion that the waveguide is very weak. On the contrary, the observed dynamic for a decaying waveguide offers an accurate way to verify the guiding characteristics. At the beginning of the process unexpected light modulations are visible on both sides of the waveguide. We believe this index perturbation located away from the waveguide is produced in the early stage of soliton formation. Indeed, when the writing beam initially diffracts, light temporary present in this region gives rise to a non-uniform space charge field prior to being trapped as a soliton in the central region. Note that these perturbations are hardly seen when the probe is ordinary polarized since the associated index change is weak. Moreover, we note that light has an even distribution during most part of the dynamic as expected from even modes excitation. However, toward the end of the erasure process we think that the transitory appearance of an asymmetric light profile is due to excitation of the second mode, which has an odd profile. It is a consequence of beam fanning [19] that tends to tilt the probe beam trajectory and leads to excitation of the second modes until the waveguide becomes singlemode. Despite these two abnormal features the overall behavior observed in the vicinity of the waveguide can be reproduced by BPM calculations. For the predicted dynamic shown in figure 6 an electro-optic $r_{33}$ coefficient around 700 pm/V has been used while other parameters are set identically than for the results in figure 2. Similarly than for the experiment with ordinary light the electro-optic coefficient has to be set to a lower value than reported in the literature (1300 pm/V) for better fit. Correlation between the two predicted electro-optic coefficient reinforces the hypothesis of a partially poled crystal.

movie-figure6.mov

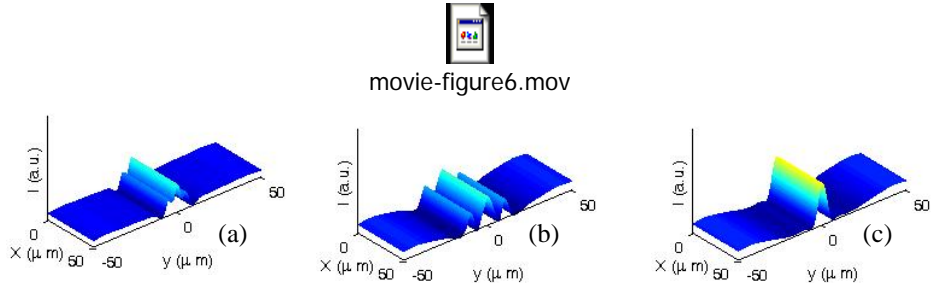

Fig. 6. Calculated dynamic of light distribution at the exit face of a decaying waveguide initially created by quasi-steady-state soliton and probe by a extraordinary polarized beam in the initial stage (a) and at two subsequent characteristic times (b, c). Movie of the entire process is attached (1.3 MB).

**Conclusion**

We have proposed and validated a simple technique to assess the guiding properties of photoinduced waveguides such as soliton induced waveguides. The technique is based on the observation of the time dependent light distribution observed at the exit face of a waveguide excited by a plane wave when its index profile is gradually evolving. Experimental demonstrations have been carried out on singlemode and multimode 1-D soliton induced waveguides in SBN crystals. The technique can be extended to test 2-D waveguides and real-time evaluation of waveguide can be envisioned to determine, for instance, the number of guided modes, the refractive index profile or the decay/formation time.

.